\documentclass[conference]{IEEEtran}
\IEEEoverridecommandlockouts
\usepackage{subcaption}
\usepackage{fancyhdr,setspace}
\usepackage{comment}
\usepackage{caption}
\usepackage{todonotes}
\usepackage{xspace}
\usepackage{bm}

\usepackage[compact]{titlesec} 
\usepackage{cite}
\usepackage{blindtext, graphicx, tabularx, multicol, lipsum, subfig, eqnarray,amsmath,chemformula,multirow,threeparttable}
\usepackage{enumitem}
\usepackage[utf8]{inputenc}
\usepackage{amssymb,amsmath}

\usepackage{float}

\usepackage[boxed]{algorithm2e}
\usepackage[noend]{algpseudocode}
\usepackage[compact]{titlesec}
\algnewcommand{\IfThenElse}[3]{
  \State \algorithmicif\ #1\ \algorithmicthen\ #2\ \algorithmicelse\ #3}
\makeatletter
\newcommand{\removelatexerror}{\let\@latex@error\@gobble}
\makeatother
\makeatletter
\renewcommand\footnoterule{%
  \kern-3\p@
  \hrule\@width\columnwidth
  \kern2.6\p@}
  \makeatother
\usepackage{letltxmacro}

\usepackage{amsthm}

\usepackage{amsmath}

\SetKwInput{KwInput}{Input}                
\SetKwInput{KwOutput}{Output}  

\usepackage{tikz}

\usepackage{capt-of}
\usepackage{amssymb} \usepackage{booktabs} 
\usepackage{pifont}

\usepackage{diagbox}
\usepackage{xcolor,etoolbox}
\usepackage{dblfloatfix}
\usepackage{color,soul}
\usepackage{placeins}
\let\mybibitem\bibitem
\renewcommand{\bibitem}[1]{%
\ifstrequal{#1}{edgeTPU}{\color{black}\mybibitem{#1}}
{\ifstrequal{#1}{xyz}{\color{blue}\mybibitem{#1}}
{\color{black}\mybibitem{#1}}}%
}

\ifCLASSINFOpdf
\else
\fi
\usepackage{xcolor}
\usepackage{comment}
\usepackage{glossaries}

\usepackage{fancyhdr}
\renewcommand{\headrulewidth}{0pt}
\begin{document}



\title{LATENT: \textbf{L}LM-\textbf{A}ugmented \textbf{T}rojan Insertion and \textbf{E}valuation Framework for Analog \textbf{N}etlist \textbf{T}opologies

}

\author{\IEEEauthorblockN{Jayeeta Chaudhuri, Arjun Chaudhuri, and Krishnendu Chakrabarty}
\IEEEauthorblockA{School of Electrical, Computer, and Energy Engineering, Arizona State University, Tempe, AZ, USA}
\\
}
\maketitle

\begin{abstract}

Analog and mixed-signal (A/MS) integrated circuits (ICs) are integral to safety-critical applications. However, the globalization and outsourcing of A/MS ICs to untrusted third-party foundries expose them to security threats, particularly analog Trojans. \textcolor{black}{Unlike digital Trojans which have been extensively
studied, analog Trojans remain largely unexplored. There has been only limited
research on their diversity and stealth in analog designs, where a
Trojan is activated only during a narrow input voltage range. Effective defense techniques require a clear understanding of the attack vectors; however, the lack of diverse analog Trojan instances limits robust advances in detection strategies. To address this gap, we present LATENT, the first large language model (LLM)-driven framework for crafting stealthy, circuit-specific analog Trojans. LATENT incorporates LLM as an autonomous agent to intelligently insert and refine Trojan components within analog designs based on iterative feedback from a detection model. This feedback loop ensures that the inserted Trojans remain stealthy while successfully evading detection. Experimental results demonstrate that our generated Trojan designs exhibit an average Trojan-activation range of $15.74$\%, ensuring they remain inactive under most operating voltages, while causing a significant performance degradation of $11.3$\% upon activation.}
\end{abstract}

\thispagestyle{fancy}
\fancyhead{}
\renewcommand{\headrulewidth}{0pt}
\fancyhf{}
\fancyfoot[C]{\thepage}


%
\IEEEpeerreviewmaketitle

\section{Introduction}

Analog and mixed-signal (A/MS) integrated circuits (ICs) are ubiquitous in many computing systems. A/MS designs are incorporated in a wide range of safety-critical applications such as automotive safety systems, military, industrial control systems, aerospace, and medical monitoring \cite{skorobogatov2012breakthrough}. These applications demand high levels of reliability, accuracy, and precision. Any compromise in the functionality of the analog IC behavior can lead to catastrophic behavior, thus compromising the safety and reliability of the systems \cite{10.1145/3342099}\cite{bhunia2014hardware}. 

As a result of globalization of the semiconductor industry, the chip design process is being outsourced to multiple foundries \cite{guin2014counterfeit}. However, this introduces security threats, commonly known as hardware Trojans \cite{elnaggar2018machine}. Hardware Trojans are malicious components that are carefully crafted to be activated under specific conditions, and remain dormant otherwise. An adversary in an untrusted third-party foundry can insert malicious logic in the design level (front-end phase) or embed fabricated trigger circuits in unused spaces of a placed-and-routed circuit \cite{7546493} \cite{5677557}. 

Several analog Trojans have been proposed in recent literature \cite{7546493} \cite{yang2017exploiting} \cite{guo2019capacitors}. In \cite{7546493}, a fabrication attack-type Trojan, A2, is developed that is triggered by a software-controlled node (referred to as \textit{victim wire}) of an OR1200 processor. The capacitive elements of the Trojan are connected to the victim wire; every time a rare operation occurs, the capacitor accumulates a charge value. When a pre-defined threshold is reached, the Trojan payload is activated.  Building on the A2 Trojan, \cite{10.1145/3489517.3530666} explored a conditionally-gated trigger that does not rely on frequently toggled nets to activate the Trojan. 

\textcolor{black}{Prior analog Trojans \cite{yang2017exploiting} \cite{guo2019capacitors} \cite{7546493}\cite{10.1145/3489517.3530666} have primarily been demonstrated on digital circuits. However, their stealth, specifically the Trojan activation range in analog circuits, remains unexplored, as analog circuits exhibit unique, non-linear behavior that are not captured by digital implementations. Moreover, these Trojans share similar structural patterns and exhibit limited stealth, making them all detectable by methods such as current-sensing \cite{9937796}, charge-depriving and capacitor-starving \cite{oriero2023demist}, watermarking \cite{10490135}, and large language model (LLM)-guided analysis \cite{chaudhuri2024spiced}. 
While existing methods \cite{9937796}\cite{10490135} \cite{chaudhuri2024spiced} can successfully identify prior analog Trojans, the lack of stealthy and diverse attack strategies prevents a through evaluation of these techniques.}
\par
To address these challenges, we introduce \textbf{LATENT}, the \textbf{first-of-its-kind} framework for generating novel analog Trojans that are specifically tailored for analog circuits. LATENT leverages an LLM as an intelligent agent that inserts Trojan components into analog designs using a feedback-driven insertion strategy. By integrating a detection tool into the reward feedback loop, the framework iteratively refines its Trojan component selection to increase the probability of evasion. Unlike previous analog Trojans that follow structurally similar patterns, LATENT generates analog Trojans that are uniquely adapted to the specific structural and behavioral characteristics of the target analog design, ensuring stealthy and functionally correct Trojan-inserted designs.

The key contributions of this paper are as follows.
\begin{itemize} [leftmargin=*,topsep=0pt]
\item We introduce LATENT, an LLM-based agentic framework that generates stealthy analog Trojan-inserted circuits.
    \item LATENT employs  a feedback-driven insertion strategy to iteratively select Trojan components and insertion points in an analog netlist to bypass detection mechanisms.
    \item LATENT integrates HSPICE for circuit simulation and SPICED \cite{chaudhuri2024spiced} to iteratively refine its attack strategy. 
    \item We evaluate the stealth and performance impact of the generated Trojans across a diverse set of open-source analog benchmark circuits.
\end{itemize}
  
The remainder of the paper is organized as follows. Section II provides an overview of analog Trojan threats, existing detection approaches, and the potential of LLM-driven methods to enhance analog security. Section III explains how LLMs can act as autonomous agents—combining iterative reasoning and decision-making. In Section IV, we introduce LATENT, an automated agentic workflow for injecting stealthy Trojans into analog circuits. Evaluation results detailing the Trojan-generation capability of LATENT, along with  comparisons among well-known analog Trojans are presented in Section V. Finally, Section VI concludes the paper.

\vspace{-0.1cm}

\section{Background and Motivation}
\vspace{-0.1cm}
\subsection{Analog Trojans and Their Impact on A/MS Design}
\vspace{-0.1cm}
Analog Trojans are malicious modifications to a circuit, either in the design phase or in the fabrication phase of the chip. These Trojans are designed to be activated under specific analog stimuli, such as voltage, temperature or process variations, and remain dormant until triggered. A Trojan consists of two primary components -- \textit{trigger} and \textit{payload}. The trigger can be either software-based or controlled externally by an input pin. The payload, upon activation, can lead to severe performance degradation or even circuit failure. A stealthy analog Trojan has the following characteristics: (1) it introduces negligible deviation in circuit performance in its dormant state, thus evading detection by traditional run-time mechanisms \cite{6691167}; (2) it is triggered when the input crosses a certain voltage threshold, and remains activated for a small input activation period; (3) it causes severe performance degradation when triggered.

In \cite{7546493}, authors propose a capacitive Trojan, namely A2, which incurs small footprint and is inserted into unused spaces of a fabricated design. Although the A2 attack was initially implemented on a digital microprocessor, recent work has shown the detrimental impact of A2 Trojan on A/MS circuit performance \cite{10490135} \cite{chaudhuri2024spiced}. In \cite{10.1145/3489517.3530666}, authors propose a glitch generator to implement the trigger circuit, which can be inserted in any node irrespective of its trigger activity. In \cite{guo2019capacitors}, several variants of charge-sharing capacitive Trojans have been implemented, that can hide in digital circuits. 
These Trojans have a small footprint and are activated only under specific conditions, making them difficult to be detected during functional verification.
\vspace{-0.1cm}
\subsection{Prior Work on Analog Trojan Detection and Prevention}
\vspace{-0.1cm}
Run-time detection of analog Trojans has been studied recently. In \cite{8383914}, authors propose an on-chip Trojan detection method that monitors a set of software-controlled nodes. If the toggling frequency of any node exceeds a threshold, a hardware interrupt is generated. Another approach relies on thermal sensors to monitor power consumption and anomalous temperature fluctuations \cite{6691167}. In \cite{9937796}, a current sensor is used to measure anomalous current spikes indicating Trojan activity.  

\textcolor{black}{While these methods have been evaluated specifically on digital circuits, their impact on analog designs is limited. Unlike digital circuits, where Trojan behavior is characterized by bit-flips, analog Trojans exploit the continuous signal variations in analog designs, making Trojan activation harder to characterize. In \cite{10490135}, authors used an analog neural twin to identify the least sensitive paths of an analog design that are most susceptible to Trojan insertion, and made them observable at circuit output through watermarks.} 

\textcolor{black}{Prior Trojan designs \cite{7546493} \cite{10.1145/3489517.3530666} rely on pre-defined structural characteristics, thereby limited in stealth and diversity. Hence, detection frameworks  [10]-[14] can effectively leverage their behavior to model and identify Trojan-induced deviations. However, these detection techniques are not robust to emerging attack vectors, where an attacker can craft stealthy and structurally diverse analog Trojans to bypass detection. }

\textcolor{black}{These limitations underscore the need for a novel approach that can automatically generate diverse and context-aware Trojans. LLMs, with their inherent ability to analyze vast contextual information and automate circuit/code generation \cite{analogcoder, verigen, qiu2024autobench, ma2024verilogreader, liu2024rtlcoder, chang2024lamagic, lu2024rtllm, chen2024llm}, offer a promising direction for designing stealthy and structurally diverse Trojan strategies that can evade existing detection frameworks. }
\vspace{-0.1cm}
\subsection{Large Language Models in Security}
\vspace{-0.1cm}
\textit{\textbf{Trojan Generation}:} LLMs have been shown to be effective for generating structural variations of digital Trojans with different trigger types \cite{bhandari2024sentaur}. Additionally, \cite{10545393} employs LLMs to insert vulnerabilities, e.g., deadlocks, into finite-state machines. In \cite{kokolakis2024harnessing}, LLMs are used to identify specific locations in the HDL code for Trojan insertion. 

However, these LLM-generated Trojans follow pre-defined structural modifications, making them more easily detectable by logic analysis and functional verification \cite{bhunia2018hardware} \cite{bazzazi2017hardware}. Moreover, they do not account for the circuit-specific nature of real-world attacks, where adversaries design Trojans that seamlessly integrate into a target circuit while evading  traditional detection mechanisms.  

\textit{\textbf{Trojan Detection}:}
LLMs have also been incorporated into bug and Trojan detection tasks \cite{10462177}. In \cite{tsai2023rtlfixer}, LLM agents are leveraged for iteratively refining erroneous Verilog code until all syntactical bugs are resolved. \cite{10545393} proposes a prompt engineering-based technique for vulnerability detection in digital circuits. Another work combines retrieval augmented generation with LLMs to systematically and iteratively patch HDL functional bugs \cite{10691874}. 
While these efforts mainly  target the detection and correction of digital Trojans, \cite{chaudhuri2024spiced} uses LLMs for identifying analog Trojans within SPICE netlists. The authors extend this work \cite{chaudhuri2024spiced} in \cite{chaudhuri2025spiced+}, where they explore Trojan mitigation strategies in A/MS designs.
\vspace{-0.1cm}
\subsection{Motivation for Proposed Solution}
\vspace{-0.1cm}
Prior work has predominantly focused on digital Trojan generation using LLM and reinforcement learning \cite{10.1145/3548606.3560690}, leaving the analog domain largely unexplored. Only a limited set of analog Trojans \cite{7546493}\cite{10.1145/3489517.3530666}\cite{9257196} have been studied. However, as analog circuits continue to be integrated more in system-on-chip designs for safety-critical applications, the threat of stealthy analog Trojans becomes increasingly severe. As noted earlier, existing defense mechanisms are not generalizable due to the limited diversity of studied Trojan instances. 

To address this limitation, our work is the first to leverage LLMs for generating analog Trojan-inserted circuits, thereby introducing a novel attack approach that has not been explored in prior research. 
This enables the automated generation of carefully crafted, syntactically and functionally correct analog Trojans. By integrating an LLM agentic workflow, our approach involves the following:

\begin{itemize}[leftmargin=*,topsep=0pt]
    \item \textbf{Natural language prompting}: The agent is guided by well-structured prompts which explain the possible modifications and structural requirements of the Trojan components, while strictly adhering to the SPICE syntax rules.
    \item \textbf{Autonomous exploration of Trojan design}: The agent autonomously proposes modifications to the design netlist to generate stealthy Trojans capable of evading detection. 
    \item \textbf{Incorporation of multiple tools}: A feedback loop incorporating HSPICE simulation and an LLM-based detection tool within the framework allows for immediate evaluation of the candidate designs, ensuring that the agent refines its strategy based on the detection outcome.
    
\end{itemize}
   \vspace{-0.1cm}
\section{Integrating Reasoning and Action via LLM Agents}
LLMs can be employed as agents that autonomously reason, make guided decisions, and perform actions in a combined manner \cite{zhao2024expel} \cite{li2024personal}.  Unlike static LLMs that generate a single response from a one-shot prompt, LLM agents operate iteratively using a self-prompting process. The agents can be customized to interact with the feedback and/or external tools to iteratively refine their strategy. In our workflow, the process begins with selecting an appropriate Trojan component and identifying node locations within the circuit-under-attack (CUA) -- the target circuit being analyzed, to generate a Trojan-inserted version of the design. Next, the framework leverages LLM-guided detection of the generated Trojan design to provide a feedback. This feedback is used to refine the component and node selection processes, and guides the agent to prioritize modifications that minimize Trojan activity while ensuring severe performance degradation upon Trojan activation.

An LLM agent follows the \textbf{Thought-Action-Observation} (TAO) framework for efficient reasoning and decision-making. The TAO framework comprises of three primary components:

\begin{itemize}[leftmargin=*,topsep=0pt]
    \item Thought ($\mathcal{T}$): The agent analyzes the CUA and formulates a strategy based on the detection feedback and the structural and behavioral characteristics of the CUA.

    \item Action ($\mathcal{A}$): In this phase, the agent selects the appropriate Trojan component and corresponding insertion nodes in the CUA based on the agent's thought.
    \item Observation ($\mathcal{O}$): The Trojan-inserted design is fed to the detection tool which provides a feedback. This feedback serves as an observation that guides the agent to refine its component and/or node selection strategy.
\end{itemize}

The TAO framework can be described as: 

\begin{equation}
\begin{aligned}
    \mathcal{T}_i &= f_1(\mathcal{O}_{i-1}, \mathcal{T}_{i-1}), \\
    \mathcal{A}_i &= f_2(\mathcal{T}_i), \\
    \mathcal{O}_i &= f_3(\mathcal{A}_i).
\end{aligned}
\end{equation}

In Equation (1), $f_1$, $f_2$, and $f_3$ represent the reasoning function, the action execution function, and the feedback function, respectively. These functions are implemented as LLM prompts that guide the agentic workflow. The framework iterates until a termination criteria is met, such as achieving a pre-defined number of Trojan components in the CUA. 

\section{LATENT: Automated Analog Trojan-Inserted Netlist Generation Framework}
\vspace{-0.1cm}

\begin{figure}
\centering
\includegraphics[width=0.48\textwidth]{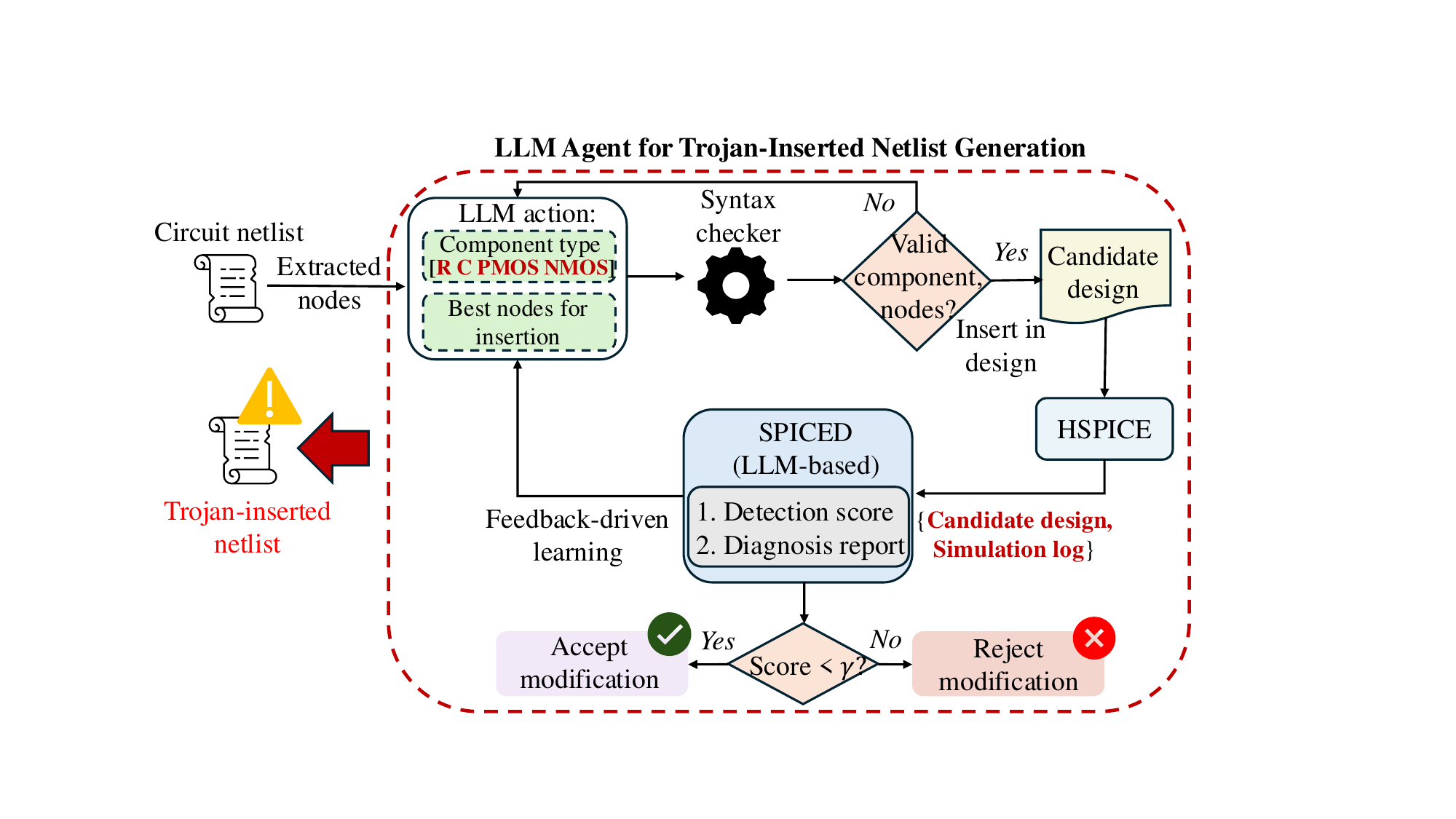}

\caption{Stealthy Trojan insertion workflow using LATENT.}
\label{latent}
\vspace{-0.6cm}
\end{figure}

In this section, we present the proposed agentic workflow, LATENT, for generating stealthy Trojan-inserted analog designs. Notably, LATENT integrates SPICED \cite{chaudhuri2024spiced}, a recently published LLM-based analog Trojan detection method. LATENT leverages this integration by guiding its agent actions based on SPICED detection feedback, thereby generating stealthy Trojans that are tailored to the specific structural and behavioral characteristics of the CUA. Fig. \ref{latent} illustrates the overall agentic workflow for Trojan-inserted netlist generation. 
\vspace{-0.6cm}

\subsection{Implementation Details} 
\vspace{-0.1cm}
\label{subsec:Implementation}

\subsubsection{Prompt Construction and LLM Agent Integration} 
\begin{figure}
\centering
\includegraphics[width=0.45\textwidth]{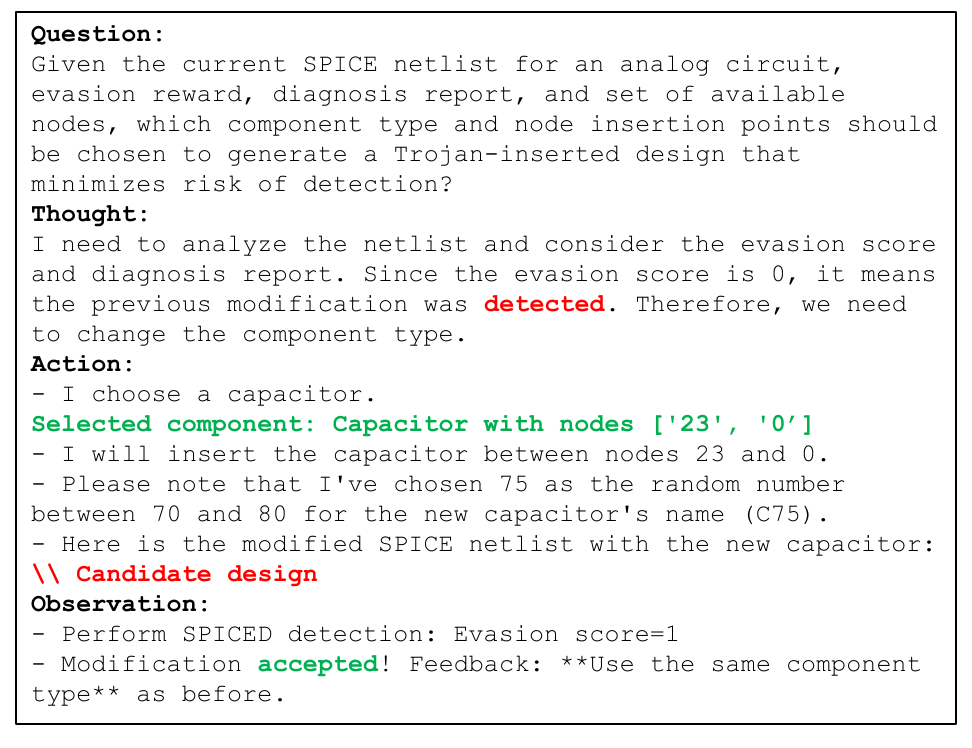}

\caption{ReACT prompts demonstrating the Thought-Action-Observation framework for Trojan insertion.}
\label{tao}
\vspace{-0.4cm}
\end{figure}

\vspace{-0.1cm}
We implement the agentic workflow using the ReAct framework within LlamaIndex\cite{yao2023react}. We construct a well-structured prompt that guides the agent to generate a sequence of TAO steps, ensuring that the agent strategically chooses the Trojan components to minimize detection. The prompt consists of the following inputs: (1) \textit{Candidate netlist} i.e., the current netlist, where the agent-generated Trojan components are inserted, (2) detection feedback score and diagnosis report from SPICED \cite{chaudhuri2024spiced}, (3) available nodes of the CUA, where the Trojan component can be inserted. The agent decides the Trojan component and the appropriate nodes of the CUA where the selected component should be inserted, based on the detection feedback. The LLM updates the candidate netlist with the new component. The ReAct steps are shown in Fig. \ref{tao}.

\textbf{Trojan insertion via feedback:} Unlike prior Trojan designs that rely on static modifications \cite{7546493} \cite{10.1145/3489517.3530666}, LATENT integrates an agent-driven feedback loop to continuously refine the Trojan modifications. LATENT achieves stealth by selecting appropriate component configurations and their insertion points. The agentic workflow ensures that LATENT-generated Trojans are significantly harder to detect compared to prior Trojans.

\textbf{Syntax checker:} Before the candidate netlist is updated, we perform a syntax check on the newly added Trojan component. We extract the agent's response, which includes the selected component type and the corresponding nodes for insertion. The response is parsed and validated to ensure that only syntactically correct Trojan components are inserted and the selected nodes exist within the available circuit nodes. If a syntax error is detected, the agent starts a new iteration. If no errors are found, the component is integrated into the netlist.

\vspace{-0.1cm}
\begin{figure}[t]
\centering
\includegraphics[width=0.5\textwidth]{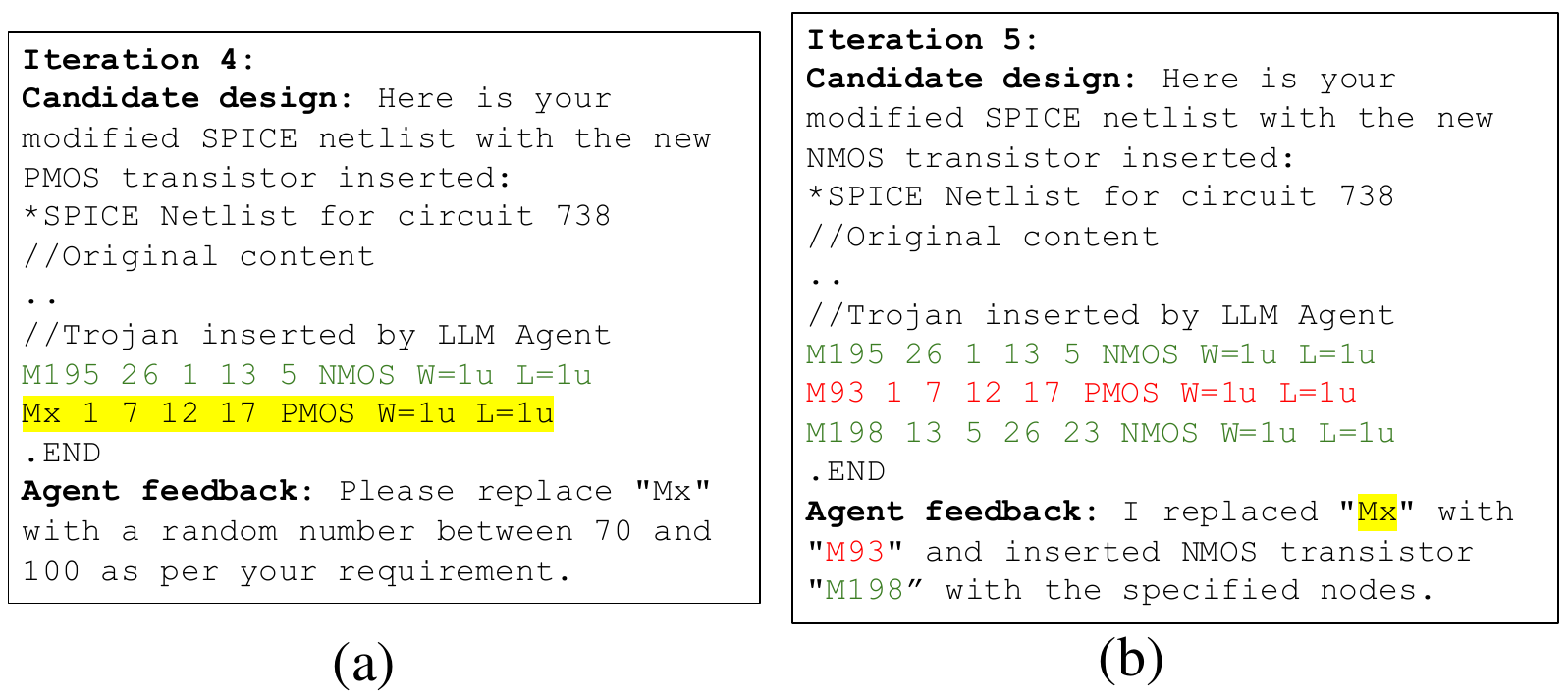}

\caption{Illustration of how the LLM agent self-corrects its previous action during the Trojan insertion process: (a) In \textbf{Iteration 4}, the agent inserts a PMOS transistor but assigns it an identifier ``Mx", which does not conform to the expected naming convention.
(b) However, in \textbf{Iteration 5}, the agent updates the identifier from ``Mx" to ``M93", following the user-defined requirement. Additionally, the agent takes a new action by inserting an NMOS transistor ``M198".}
\label{candidate}
\vspace{-0.5cm}
\end{figure}

\subsubsection{SPICED Prompting}
\vspace{-0.1cm}
Since the SPICED \cite{chaudhuri2024spiced} codebase is not yet publicly available, we re-implemented the framework by replicating the supervised-learning approach and few-shot prompting techniques. As \cite{chaudhuri2024spiced} is explicitly provided with few-shot examples that contain only capacitive and transistor-based Trojans, the detection model will likely be biased against identifying resistor-based Trojans. One approach to solve this is to include resistor-based Trojans in the few-shot examples. \textcolor{black}{However, to the best of our knowledge, no resistor-based analog Trojans have been described in prior literature, making it infeasible to provide such examples.} To ensure a realistic attack scenario, we instead add explicit instructions in the prompt to ensure that resistor-induced anomalies are considered by the model.  
\vspace{-0.1cm}
\subsection{Component Selection and Randomized Insertion}
\vspace{-0.1cm}
A critical step in developing a Trojan-inserted design is selecting the appropriate component type and node insertion points within the design. \textcolor{black}{As explained in \cite{7546493} \cite{10.1145/3489517.3530666} \cite{guo2019capacitors}, analog Trojans are typically capacitor, NMOS, or PMOS-based. Our approach expands beyond these designs by also considering resistor-induced anomalies.} We employ a well-defined, structured prompting approach to guide the agent to select components from a pre-defined set: resistors, capacitors, NMOS, PMOS. Next, the agent identifies viable nodes within the netlist for inserting the Trojan components. Based on the selected component type, the framework samples the required number of nodes (two for passive elements, three for MOSFETs) from the nodes present in the CUA. For example, the agent chooses a capacitive Trojan component in the format: \textit{C$<$x$>$ n1 n2 val}, where \textit{$<$x$>$} is an automatically chosen random identifier, \textit{n1} and \textit{n2} are nodes selected from the CUA, and \textit{val} is the randomly chosen capacitor value to be consistent with the parameter setup in prior work \cite{7546493}. The LLM agent updates the candidate netlist by appending a SPICE-formatted line corresponding to the Trojan component. \textcolor{black}{Note that 
unlike a sequential insertion method, the LLM-based agent employs randomization to distribute the Trojan insertions non-contiguously throughout the netlist, making the modifications harder to detect via pattern-based analysis \cite{6992221}}.  

The LLM agent exhibits inherent intelligence in detecting SPICE syntax inconsistencies. Fig. \ref{candidate} demonstrates one such example where the agent updates the PMOS identifier incorrectly during the Trojan-insertion process. However, in subsequent iterations, it refines the PMOS identifier. This showcases the LLM's ability to recognize inconsistencies and dynamically refine them to align with the prompt requirements.

\vspace{-0.1cm}
\subsection{Iterative Detection Feedback for Stealthy Trojan Insertion}
\vspace{-0.1cm}
After the agent either inserts a Trojan component or reverts its previous action, the candidate netlist is simulated using HSPICE to produce a simulation log, denoted as $Sim_{netlist}$. 

The candidate netlist and $Sim_{netlist}$, which records both voltage and current behavior of circuit nodes, are fed as inputs to SPICED for analysis. SPICED generates a set of suspect lines, $L_{sus}$ and a detailed diagnosis report. From the report, we extract two key pieces of information: \textbf{(1)} $L_{sus}$, and 
\textbf{(2)} detailed analysis of why these lines were flagged, including node-level anomalies in voltage/current behavior under certain input test conditions. We use information \textbf{(1)} to compute \textbf{(3)} the evasion reward $R_{evade}$, which is formulated as: $R_{\text{evade}}=\frac{L_{\text{agent}} \cap L_{\text{sus}}}{L_{\text{agent}}}\times 100\%$, where $L_{agent}$ represents the Trojan components inserted by the agent.
\vspace{-0.1cm}

\subsection{Stealthy Trojan Insertion Strategy}
\vspace{-0.1cm}
The agent autonomously analyzes feedback \textbf{(2)} and \textbf{(3)} to adapt a strategic Trojan-insertion approach as follows:

\begin{itemize}[leftmargin=*,topsep=0pt]
    \item \textbf{$R_{evade}=100\%$}: This indicates that the added Trojan component in the candidate design evades detection by SPICED. There is no new insight to guide the agent for a change in its action in the next iteration. Therefore, the agent \textit{retains the component type} in the next iteration.
    \item Using the diagnosis report (applicable if $R_{evade}<100\%$): If $R_{evade}<100\%$, LATENT leverages SPICED diagnosis report to explore a new component type/node insertion points. This approach ensures that the agent \emph{reverts} to the previous netlist configuration that was previously undetected.
\end{itemize}
\begin{figure}
\centering
\includegraphics[width=0.4\textwidth]{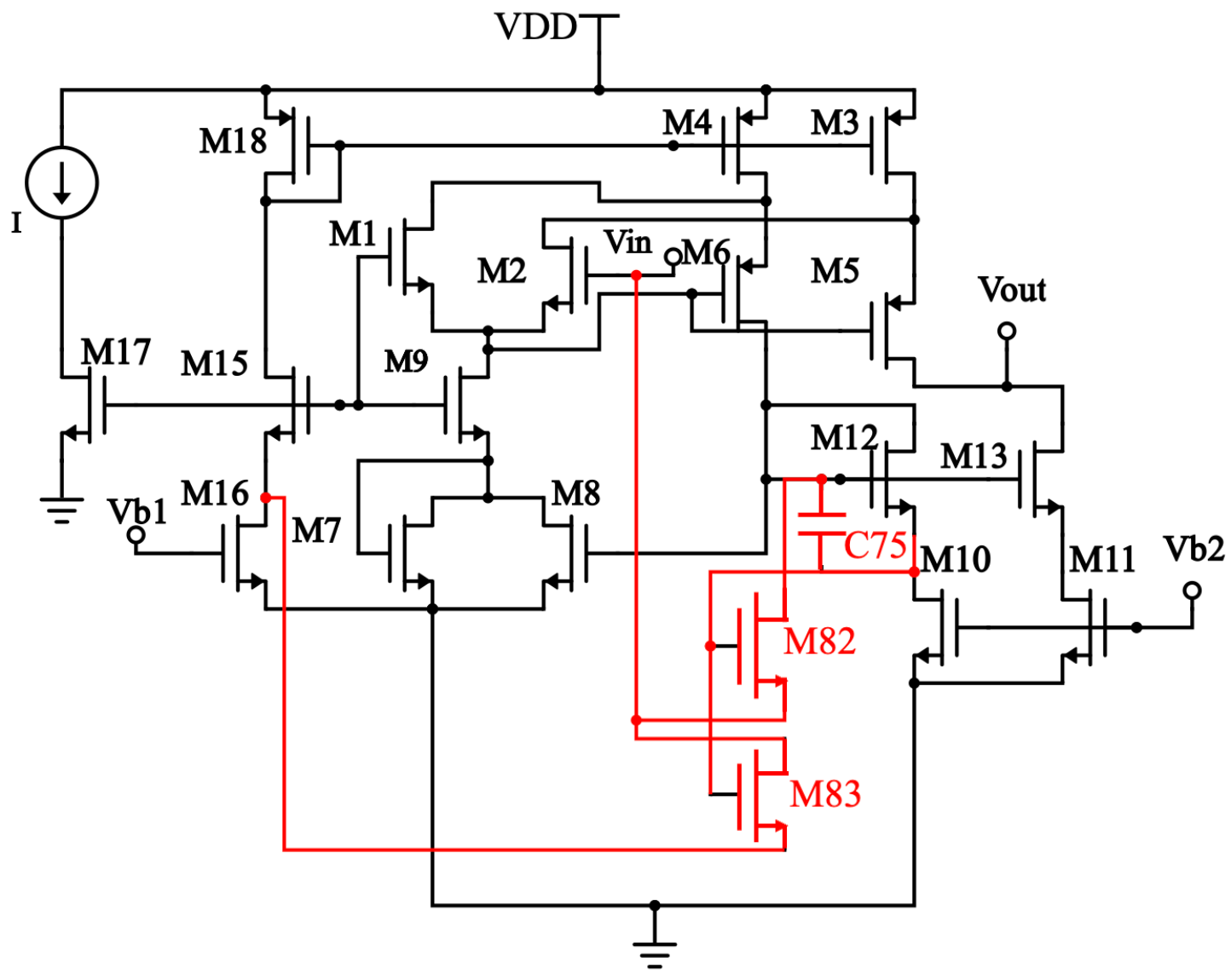}

\caption{A Trojan-inserted design corresponding to `642' of AMSNet \cite{amsnet}, generated by LATENT. The red-highlighted
components, including a capacitor and two NMOS transistors,
represent the Trojan elements inserted by the agent.}
\label{troj_design}
\vspace{-0.6cm}
\end{figure}
\begin{table*}[t]
\centering
\caption{Stealth and effectiveness of LATENT-generated Trojans for several analog designs.}
\fontsize{6.9}{6.9}\selectfont

 \begin{tabular}{|c|c|c|c|c|c|c|c|c|c|c|} 
   \hline

Case&Netlist&\multicolumn{2}{c|}{Trojan Type}&\multirow{2}{*}{$n_{it}$}&\multicolumn{2}{c|}{$L_T$}&\multicolumn{2}{c|}{$R_{evade}$ (\%)}&\multicolumn{2}{c|}{Node Impact Score (\%)}\\ 
\cline{3-4}
\cline{6-11}
&&One-shot&LATENT&&One-shot&LATENT&One-shot&LATENT&One-shot&LATENT \\
\cline{1-11}
1&738&C&C+NMOS&8&1&4&0&100&11.1&26 \\
2&642&C+NMOS&C+NMOS&9&2&3&0&100&22.2&27.7 \\
3&669&C+NMOS&R+NMOS&9&5&3&20&100&43.7&18.75 \\
4&671&C+NMOS&R+C+NMOS&8&4&6&25&100&29.4&23.5 \\
5&672&C+PMOS+NMOS&C+NMOS&7&6&3&33.3&66.6&19&33.3 \\
6&673&C&R+NMOS&6&3&3&0&100&9&22.7 \\
&&&+PMOS&&&&&&& \\
7&681&NMOS+PMOS&C+R+NMOS&9&2&4&50&75&40&40 \\
8&684&C+PMOS+NMOS&C+NMOS+PMOS&9&7&4&42.8&100&40&26.6 \\

9&693&NMOS&R+NMOS&8&3&4&33.3&100&10.5&19 \\
10&705&R+C&C+NMOS&7&5&3&60&100&19&15.7\\
11&716&C+NMOS&C+NMOS&7&4&4&0&75&16.6&19 \\
12&755&NMOS+PMOS&R+NMOS+PMOS&6&6&4&50&100&28.5&26.3\\

13&OPAMP&C+NMOS&C+NMOS&8&4&4&25&100&33.3&35.7\\
14&LDO&R+C+NMOS&C+NMOS+PMOS&12&6&5&16.6&100&0.8&0.7 \\

15&Bandgap &NMOS&C+R+NMOS&9&2&3&0&100&2.2&2.6\\
&filter &&&&&&&&&\\
\hline
\hline
\multicolumn{7}{|c|}{\textbf{Average}}&23.7&\textbf{94.4}&21.6&\textbf{22.5} \\

\hline

\end{tabular}
 \begin{tablenotes}
\item Trojan type is categorized as C: Capacitive, R: Resistive, NMOS/PMOS: Transistors. $n_{it}$ indicates the minimum number of iterations required by LATENT to converge to a stealthy Trojan-inserted design which achieves either $R_{evade}=100\%$ for 3 consecutive iterations or reaches $L_{max}$.
\end{tablenotes}
 \vspace{-0.4cm}
 \label{trojresults}

\end{table*}

Based on the above steps, LATENT outputs a Trojan-inserted netlist that: (1) includes strategically placed Trojan components, (2) induces performance degradation on Trojan activation, and (3) evades SPICED detection, ensuring stealth. 

Fig. \ref{troj_design} shows the agent-guided Trojan-inserted design of circuit `642' \cite{amsnet}, where the selected Trojan components are strategically placed based on SPICED feedback.
\vspace{-0.2cm}

\subsection{Determining the Upper Bound of Trojan Components} 

\vspace{-0.1cm}
We address two key challenges in selecting the upper bound of Trojan components:
\begin{itemize}[leftmargin=*,topsep=0pt] \item \textit{Over-insertion}: Inserting a large number of Trojan components (e.g., transistors) in a design can lead to increased area and power overheads, and expands the Trojan activation range, making detection more likely \cite{chaudhuri2024spiced}. \item \textit{Under-insertion}: Fewer Trojan components (e.g., a single capacitor) may not induce significant deviation in system performance, making the Trojan ineffective. \end{itemize}

\textit{Solution:} To ensure a balance between stealth and impact, we consider a heuristic approach in determining an upper bound on the number of inserted Trojans. The upper bound $L_{max}$ is formulated as: $L_{max}=\alpha N$, where $N$ is the total number of nodes in the CUA, and $\alpha$ is an user-configurable parameter that can be tuned based on computational constraints. 

\textbf{Termination criteria:} The iteration process stops when one of the following conditions is satisfied: 
\begin{enumerate}[leftmargin=*,topsep=0pt]
    \item $R_{evade}=100\%$ for $T$ consecutive iterations: $T=1$ is not optimal, as SPICED might have false positives in a single run \cite{chaudhuri2024spiced}. Therefore, we require $R_{evade}=100\%$ for multiple consecutive iterations, thus ensuring that the inserted Trojan components consistently evade detection. 
    
    \item Upper bound $L_{max}$ reached: If the number of inserted Trojan components reaches $L_{max}$, the process terminates. 
\end{enumerate}

\vspace{-0.1cm}




\section{Experimental Results}

\vspace{-0.1cm}
\subsection{Experimental Setup}
\vspace{-0.1cm}
We implement LATENT using Python-3.12. We use the \textit{GPT-4o-mini} API model for LATENT and SPICED implementations. We use the default temperature setting of 0.3. We simulate the candidate netlists using HSPICE. We evaluate LATENT on (i) the open-source analog benchmark repository AMSNet \cite{amsnet}, specifically choosing the circuits with the highest nodes (in the range 20-25), (ii) bandgap filter (265 nodes), and (iii) LDO (1655 nodes). Based on experimental observations, we set the termination thresholds (described in Section IV.E) to $T=3$, with
$\alpha$ set to 0.6. \textcolor{black}{$T=3$ ensures that evasion is consistently achieved while preventing early termination of the framework. Similarly, setting $\alpha=0.6$ prevents excessive circuit modifications that could make the Trojans easily detectable. }The experiments are carried out on a 3 GHz AMD EPYC 7313 CPU with 2 TB of RAM. 
\vspace{-0.1cm}
\subsection{Evaluation Metrics}
\vspace{-0.1cm}
We use the following performance metrics to determine the effectiveness of LATENT:
\begin{itemize}
    \item Node impact score (\%): $\frac{n_{tr}}{N}\times100$, where $n_{tr}$ is the number of Trojan-impacted nodes.
    \item Activation range (\%): $\frac{\textit{Trigger activation inputs}}{\textit{Total input space}}\times100$. 
    \item $R_{evade}$: Percentage of inserted Trojans that evade detection by SPICED.  
    \item $n_{it}$: Number of iterations required in the agentic workflow to achieve a feasible Trojan-inserted design.
    \item $L_T$: Number of Trojan components inserted in the CUA.
\end{itemize}
\vspace{-0.1cm}
\subsection{Stealthy Trojan-Inserted Designs} 
\vspace{-0.1cm}

\begin{figure}
\centering
\includegraphics[width=0.36\textwidth]{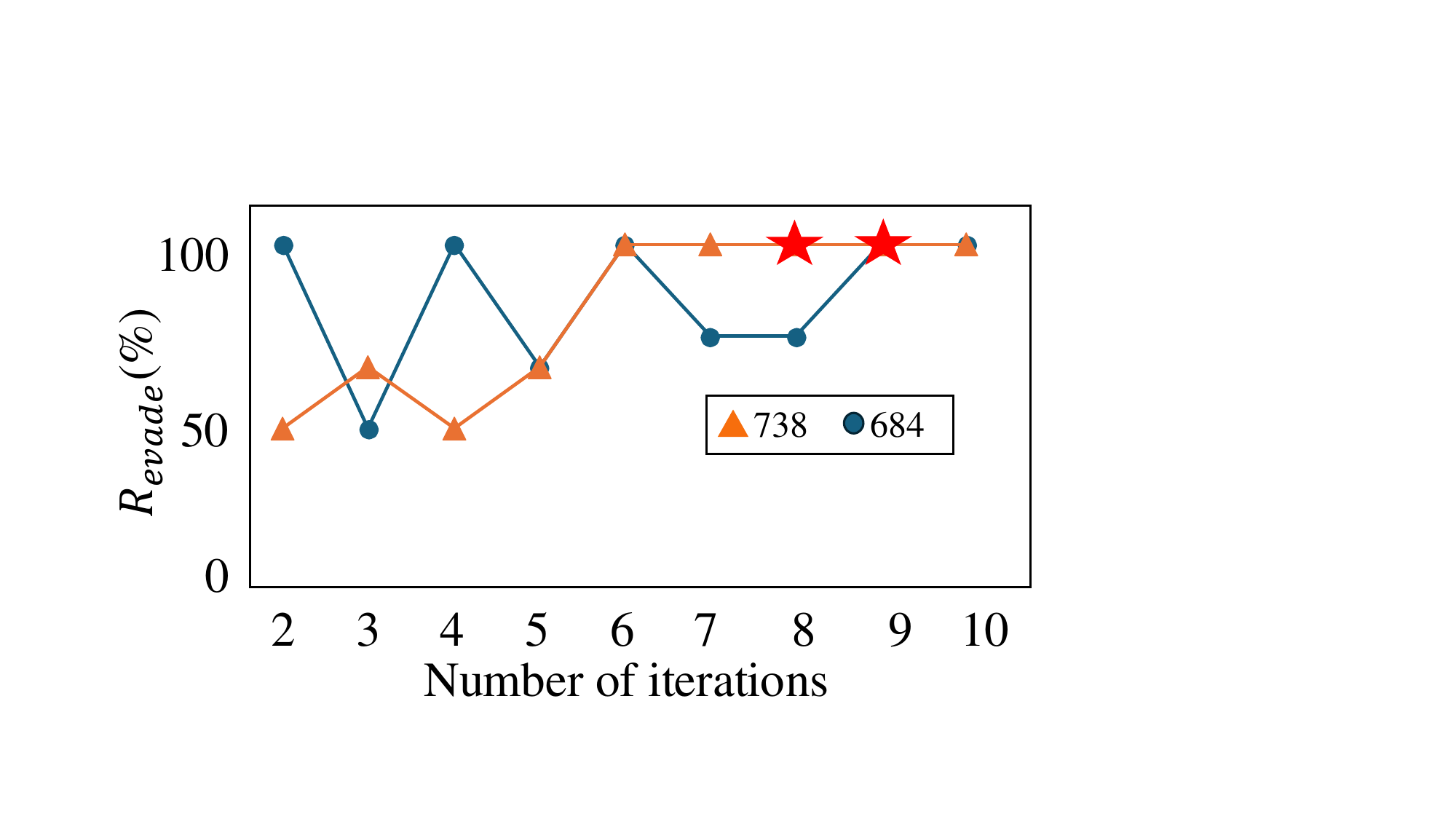}

\caption{$R_{evade}$ convergence across iterations for netlists from AMSNet \cite{amsnet} (\textcolor{red}{$\bigstar$} indicates the minimum number of iterations required for convergence). }
\label{example}
\vspace{-0.2cm}
\end{figure}

Table \ref{trojresults} reports the impact of the LATENT-generated Trojan on the activation range and detection evasion. We use one-shot prompting approach as the baseline, where the LLM generates a Trojan-inserted netlist in a single attempt, without any iterative refinement or feedback loop. From Table \ref{trojresults}, we observe that Trojans inserted via one-shot prompting are easily detectable by SPICED. In contrast, LATENT follows a strategic, feedback-driven approach to generate more stealthy Trojans. Moreover, Fig. \ref{example} illustrates how $R_{evade}$ evolves with each iteration as the agent strategically inserts a Trojan component or reverts its action, leveraging SPICED feedback. While Fig. \ref{example} presents results for netlists `738' and `684', a comprehensive analysis of $R_{evade}$ for all the evaluated netlists is provided in the appendix (Fig. \ref{merged}). 

\vspace{-0.1cm}
\subsection{Resource and Time Overheads}
\vspace{-0.1cm}
For OPAMP and AMSNet \cite{amsnet} circuits, the input token size of the agent and SPICED reached 4600 and 18000, respectively. Larger circuits such as the bandgap filter and the LDO significantly increased the token requirements, with the agent and SPICED requiring 33500 and 108000 tokens, respectively. Since GPT-4o-mini's input cost is around \$0.15 for 1M tokens and output cost is \$0.6 per 1M tokens, LATENT offers a cost-effective solution for Trojan-inserted netlist generation \cite{openai}.
\begin{table}
\centering
\fontsize{7}{7}\selectfont
\caption{\textcolor{black}{Breakdown of average runtime overheads of LATENT (the reported times for thought-action, SPICE simulation, and SPICED analysis correspond to a single iteration).}}

    \label{compare}
    \begin{tabular}{|c|c| c|c|c|c|} 
    \hline
Circuit&\multicolumn{3}{c|}{LATENT (LLM agent)}&\\
\cline{2-4}
&Thought-Action&SPICE&SPICED&Total \\
&(s)&simulation (s)&analysis (s)&time (s)\\
\hline
AMSNet/OPAMP&2.5&0.3&17.4&80.8 \\
Bandgap filter&2.8&0.8&18.1&130.2 \\
LDO&2.8&1.1&18.6&157.5 \\

\hline
\end{tabular}

\label{time}

\vspace{-0.4cm}
\end{table}
Table \ref{time} reports  the runtime overhead of LATENT. Notably, the runtime can be significantly reduced if GPUs are used. Since LATENT is platform-agnostic, it can be adapted for GPUs or TPUs based on user requirements.
\vspace{-0.1cm}
\subsection{Comparison with State-of-the-art}
\vspace{-0.1cm}
We evaluate the LATENT-generated Trojans against baseline Trojans, A2 and DELTA. For a fair comparison, we insert A2 and DELTA at random nodes per netlist and compute their average impact on the output performance deviation $\Delta P$, computed as: $\Delta P=\frac{|V_o-V_{max}|} {V_o}\times 100\%$, where $V_{max}$ is the output voltage that shows the maximum deviation in the Trojan-inserted design, and $V_o$ is the corresponding output voltage of the Trojan-free design, for the same input. The area overhead $\Delta A$ is computed as: $\Delta A=\frac{\sum_{i=1}^{N_T} W_i\times L_i}{A}\times 100\%$, where $N_T$ is the number of transistor-based Trojan components, $W(L)$ indicates the width (length) of the transistor, and $A$ refers to the original area of CUA. Table \ref{area} presents a detailed comparison; we observe that the LATENT-generated Trojans have a low activation range and minimal area overhead by strategically selecting components and insertion points based on the specific CUA and SPICED feedback, unlike A2 and DELTA, which have fixed structural patterns.
 \vspace{-0.1cm}
\begin{table}
\centering
\caption{Comparison of performance and area overheads with prior analog Trojans.}
\label{tab:trojan_comparison}
\fontsize{6.5}{6.5}\selectfont
\begin{tabular}{|c|c|c|c|c|c|c|}
 \hline
Netlist & \multicolumn{3}{c|}{$\Delta P$ (Activation Range) (\%)} & \multicolumn{3}{c|}{$\Delta A$ (\%)}   \\

\cline{2-7}
& A2 & DELTA &  \textbf{LATENT} &A2 & DELTA &\textbf{LATENT} \\
\cline{1-7}

738 & 6.8 (46.1)&9.6 (46.1) &13.3 (11.1) & 20&24&    8.7 \\
642 & 20.1 (57.6)&8.5 (46.1) &9.3 (11.1) &22.2 &26.6 &  10.6  \\
669 & 7.2 (23)&9.5 (30.7) &10.8 (10) &15.3 &18.4&  11.5   \\
671 & 9.1 (57.6)&5.3 (23)&12.2 (23) &20 &25& 10   \\
672 &5.8 (23) &9.1 (46.1)&2.5 (23) & 17.6&23 &  5.8  \\
673 &23 (30.7) &15.7 (46.1) &17.4 (11.5) &15.7 &21&  7.8  \\
681 &12 (23) &21.5 (57.6)& 15 (19.2)& 37.5 &48 & 6.2   \\
684 &28.7 (46.1) &21 (46.1)&13.8 (11.5) &25 & 27.6& 9  \\
693 &7.7 (23) &1.7(46.1) &12.7 (11.5) &33.3 & 48& 8.3   \\
705 &3.5 (30.7)  &4.4 (57.6) &9.8 (23) &16.6&20&  6.2 \\
716 &6.1 (57.6) &6.3 (57.6) &8.2 (11.5) &16.6&20&   8.3  \\
755 &13.2 (30.7)  &8.3 (46.1)&14.6 (11.5) &23& 25& 8.3    \\
OPAMP &8.2 (23)&3.1 (30.7)&3.7 (19.2) &17.6&23 &  5.8   \\
LDO &8.2 (19.2) &10.5 (23) &19.1 (19.2)&  0.8& 0.86&  0.6\\
Bandgap &\multirow{2}{*}{1.8 (23)}& \multirow{2}{*}{5.7 (46.1)}&\multirow{2}{*}{6.3 (19.2)}  &\multirow{2}{*}{6.8} &\multirow{2}{*}{9} & \multirow{2}{*}{3.4}  \\
filter &   & & & & &   \\
\hline
\hline
\multicolumn{1}{|c|}{\textbf{Average}}&10.7 (34.3)&9.1 (43.2)&\textbf{11.3 (15.7)}&19.2&23.9&\textbf{7.4} \\

\hline
\end{tabular}
\vspace{-0.4cm}
\label{area}
\end{table}

\subsection{Discussion}
\vspace{-0.1cm}
Our work demonstrates promising results across the evaluated designs, including smaller circuits ($\sim$20 nodes). Note that these circuits are often submodules within larger, complex A/MS systems. Even if a submodule exhibits over-insertion of Trojan components compared to its size, its effects may remain concealed within the overall system. For example, a capacitive Trojan in an OPAMP may have a large activation range in isolation but remain undetectable within an ADC.

Currently, no well-documented method exists for analyzing the AC and small-signal characteristics of analog Trojans. Since SPICED \cite{chaudhuri2024spiced} detects Trojans based on their DC characteristics, we have specifically generated DC-activated Trojans in our work. For future work, we will explore AC and small-signal analysis to develop a dedicated detection model, which will subsequently enable our agentic framework to diversify Trojan generation in these domains. We will also explore how parametric changes such as W/L, capacitive coupling etc. can further diversify Trojan strategies.

\section{Conclusion}
LATENT leverages LLM agents for feedback-guided Trojan insertion in analog designs. The generated Trojans incur low area overhead while achieving significant performance degradation upon activation. These findings highlight the potential of agent-driven frameworks for analog design security.


\newpage

\bibliographystyle{myIEEEtran}

\bibliography{references}
\newpage
\appendix

\begin{figure}[h]
\centering
\includegraphics[width=\columnwidth]{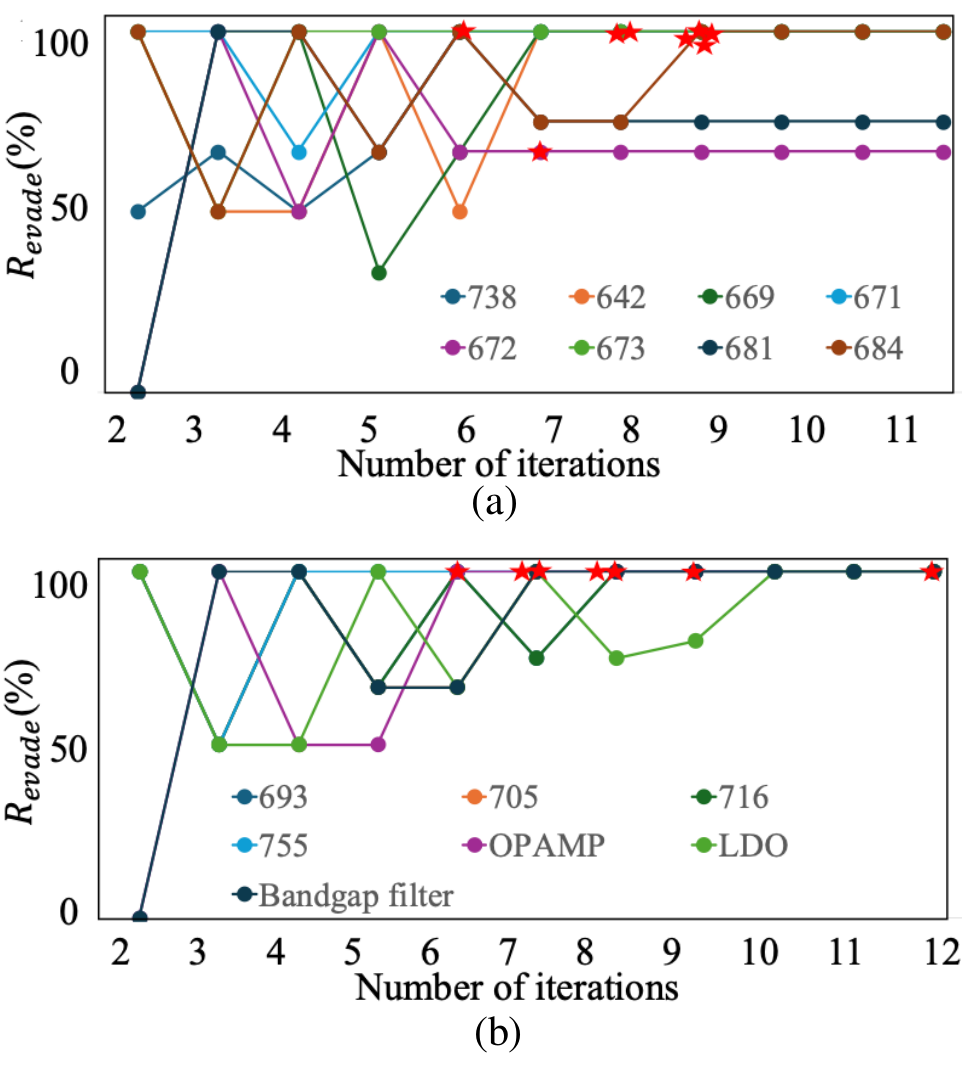}

\caption{$R_{evade}$ convergence across iterations for all the evaluated netlists  (\textcolor{red}{$\bigstar$} indicates the minimum number of iterations required for convergence).}
\label{merged}
\vspace{-0.2cm}
\end{figure}

\textbf{$R_{evade}$ vs. iterations:} Fig. \ref{merged} illustrates the convergence trend of $R_{evade}$ for the evaluated netlists.

\textbf{Diversity of Trojan components in LATENT-generated Trojan designs:} LATENT leverages a feedback-based approach to iteratively refine its selection of Trojan component types and their corresponding node locations for insertion. Unlike one-shot prompting approach, which selects and inserts Trojan components based entirely on the LLM's initial response without relying on any feedback, LATENT continuously updates its actions based on SPICED feedback. This feedback-driven refinement allows LATENT to generate a diverse set of Trojan components that are uniquely tailored to each CUA based on its specific structural and behavioral characteristics. 

\end{document}